\documentclass{elsart}
\usepackage{graphicx,amssymb,amsmath}
\usepackage[a4paper]{geometry}
\usepackage{ulem}
\usepackage{color}
\usepackage{rotating}
\usepackage{epsfig}
\usepackage{wasysym}
\usepackage{bbm}
\usepackage{multirow}
\usepackage{subcaption}

\newcommand{\s}{{\sigma}}

\def\be{\begin{eqnarray}}
\def\ee{\end{eqnarray}}

\newcommand{\nn}{\nonumber\\}

\newcommand{\Eq}[1]{Eq.~(\ref{#1})}
\newcommand{\<}{\langle}
\renewcommand{\>}{\rangle}
\newcommand{\Tr}{{\rm Tr}}

\newcommand{\p}{\partial}

\newcommand{\ra}{\rightarrow}

\newcommand{\Fig}[1]{Fig.~\ref{#1}}

\newcommand \ti[1]{}
\newcommand \jn[1]{{#1}}

\begin{document}

\begin{frontmatter}

\title{Critical theories of phase transition between symmetry protected topological states and their relation to the gapless boundary theories
}
\author{Xie Chen$^{1}$,Fa Wang$^{2}$,Yuan-Ming Lu$^{3}$,Dung-Hai Lee$^{1,3}$}
\address{$^{1}$Department of Physics, University of California at Berkeley, Berkeley, CA
94720, USA;\\
$^{2}$International Center for Quantum Materials and School of Physics,
Peking University, Beijing 100871, China;\\
$^{3}$Materials Sciences Division, Lawrence Berkeley National Laboratory,
Berkeley, CA 94720, USA}

\begin{abstract}
Symmetry protected topological states (SPTs) have the same symmetry and the phase transition between them are beyond Landau's symmetry breaking formalism. In this paper we study (1) the critical theory of phase transition between
trivial and non-trivial SPTs, and (2) the
relation between such critical theory and the gapless boundary theory of SPTs.
Based on examples of SO(3) and SU(2) SPTs, we propose that under appropriate boundary condition the critical theory contains
the delocalized version of the boundary excitations. In addition, we prove that the boundary theory is
the critical theory spatially confined  between two SPTs. We expect these conclusions to hold
in general and, in particular, for discrete symmetry groups as well.
\end{abstract}



\end{frontmatter}

\newpage

\section{Introduction}
In the past year, much progress has been made in identifying a new (interacting) class of states of matter -- the so-called ``symmetry protected topological states'' (SPTs)\cite{xie}. Topological insulators and superconductors are nice examples of SPTs. SPTs possess energy gaps in the bulk of the system and do not break any symmetry of the Hamiltonian. They have symmetry protected gapless boundary excitations. In Ref.\cite{xie} it is argued that different SPTs are labeled by the elements of the cohomology group associated with the symmetry of the system. There are a number of outstanding issues that need to be understood in this new field. The subject of this paper concerns the critical theory of phase transition between these states.

In traditional statistical physics phases of matter are characterized by symmetries, and phase transitions mark the change in symmetry\cite{Landau 1937}. They are described by the theory of Ginzburg-Landau\cite{GL} and Wilson \cite{Wilson}. In contrast, different SPTs have the same symmetry but differ in the way symmetry is represented by their boundary excitations. A natural question is what differentiates the critical points between different SPTs and the usual Ginzburg-Landau-Wilson critical points. A less obvious but equally natural question is what is the relation between the theories for such critical points and the theories of gapless excitations at the boundary of SPTs.

On the surface, it seems unjustified to expect a relationship between a bulk critical theory and a boundary theory, since they live in different space dimensions. The reason we expect such a relationship is the following physical picture for the gapless excitations at the boundary of non-trivial SPTs. Vacuum is a trivial SPT, it can not be connected to the nontrivial SPT in the bulk without crossing a phase transition. Now consider varying a parameter smoothly from inside to outside the sample so that the system evolves from a non-trivial SPT to the trivial SPT. The necessary phase transition then gives rise to the gapless excitations near the boundary. According to this picture {\it the gapless boundary of a non-trivial SPT is the bulk critical theory spatially confined between two different SPTs.}

For free fermion SPTs\cite{kitaev,ryu}, it is a simple exercise to see that the above picture is indeed true. In brief, the trivial and non-trivial SPTs are produced by adding ``mass'' of different signs to the critical theory. Boundary corresponds to the mass domain wall, in the presence of which only momentum parallel to the interface is conserved. The gap node(s) in the boundary theory can be obtained from those of the bulk critical theory with the component of momentum perpendicular to the interface deleted. It is much less clear how the picture works out for interacting SPTs.

In this paper, we study two examples of SPT phase transitions in interacting systems, that between the $SO(3)$ SPTs in one dimension (with $Z_2$ classification) and the one between $SU(2)$ SPTs in two dimension (with $Z$ classification). We see from these examples that (1) When subjected to appropriate boundary conditions the critical theory possesses the delocalized version of edge excitations (2) The edge theory is the bulk critical theory spatially confined between two SPTs, as is the case for free fermion systems. We expect these results to apply to general critical theories of phase transitions between SPTs, in any dimension and with any symmetry.

The paper is organized as follows. In section \ref{SO(3)_1d}, we study the phase transition between the $1+1d$ SPTs with $SO(3)$ symmetry and its relation to the degenerate edge state. Section \ref{SU(2)_2d} is then devoted to $2+1d$ SPTs with $SU(2)$ symmetry where the critical theory for phase transition is connected to the gapless edge states through the coupled wire construction of the SPTs. While in section \ref{SU(2)_2d} we focus mainly on the phase transition between vacuum and the first nontrivial SPT in the $Z$ classification, in section \ref{stacked_layer} we discuss how the critical theory between vacuum and other nontrivial SPTs can be obtained starting from the discussion in section \ref{SU(2)_2d}. Finally in section \ref{conclusion}, we summarize what we have learned from these examples and discuss open questions.

\section{The phase transition between SO(3) SPTs in 1+1 dimensions}
\label{SO(3)_1d}

\begin{figure}[!htb]
\begin{center}
\includegraphics[scale=0.35]
{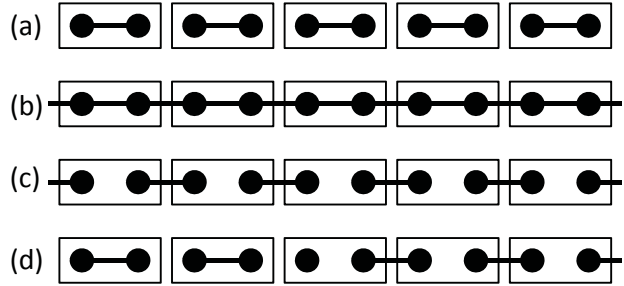}\hspace{0.1in}
\caption{(color on-line) Spin 1/2 chain with alternating nearest neighbor antiferromagnetic bond strength. The rectangular box encloses the unit cell. (a) The dimmer phase, (b) the critical point, (c) the Haldane phase, (d) the domain wall configuration between (a) and (c).\label{fig1}}
\end{center}
\end{figure}

In one space and one time dimension there are two distinct SPTs protected by SO(3) symmetry\cite{Pollmann1, Pollmann2, Chen1d, Schuch1d}. A possible realization of these two phases is given by the two dimer states of a {\it spin 1/2} nearest-neighbor antiferromagnetic chain with alternating antiferromagnetic bond strength shown in panel (a) and (c) of \Fig{fig1}.

Why do we regard the two phases in question as SO(3), not SU(2), SPTs?  First, we can define the spatial unit cell (e.g. the boxes in \Fig{fig1}) so that the total spin in each cell is integral, hence carrying the faithful (``linear'') representation of SO(3). Ultimately what determines whether a phase has stable gapless edge states is the following rule: what degrees of freedom and what kind of perturbations are we allowed to add locally, and hence to the edge. If we regard the ``on site'' (or ``on cell'') symmetry group as SO(3), then only integral spins are allowed to be added to the edge. Since the edges in \Fig{fig1}(c) have half integer spins, adding integer spins can not completely gap them out. On the other hand if we regard the on site symmetry group as SU(2), half integer spins can be added. In that case the edge spins in \Fig{fig1}(c) are not ``symmetry protected''. Therefore, \Fig{fig1}(c) defines a nontrivial SPT phase only with $SO(3)$ symmetry and with each unit cell containing two spin 1/2's. Defining the unit cell also sets the rule for spatial truncation when a chain is cut open. {\it A finite spin chain must include an integer number of unit cells.} When we truncate a spin chain we are not allowed to cut through the unit cells.

Another question that requires immediate attention is the following. In \Fig{fig1} if we move all unit cells by one lattice spacing (between two spin 1/2's not two unit cells) and complete the chain by adding the missing part of the unit cells at the two ends, then the role of panel (a) and (c) will be exchanged. Hence there seems to be ambiguity in which phase should be called trivial/non-trivial. Indeed, trivialness and non-trivialness only have relative meaning. The best way to tell whether two phases are in the same topological class or not is to make an interface between them and see whether there is gapless interface states (see \Fig{fig1}(d)). The non-trivial SPT in \Fig{fig1}(c) can be adiabatically transformed to the Haldane phase\cite{haldane,Hida,Kohmoto} by turning on an intra unit cell ferromagnetic interaction. There is no phase transition as we turn up the ratio between the ferromagnetic to antiferromagnetic interactions.

\subsection{The critical theory}
The critical point between the trivial and the non trivial SPTs in \Fig{fig1} is the gapless Heisenberg S=1/2 chain (see \Fig{fig1}(b)).
 It's effective field theory is the level-1 SU(2) (SU(2)$_1$) Wess-Zumino-Witten (WZW) theory in 1+1 dimensions\cite{Affleck-Haldane}:
\be
&&S_{\rm critical}=\frac{1}{2\gamma_1}\int dxdt \sum_{j=x,t}\Tr(\partial_j g^{-1}\partial_j g)+S_{\rm{WZW}}\nn&&S_{\rm{WZW}}=\frac{i}{2\pi}\int dxdt\int_0^1du\Tr[(g^{-1}\p_u g)(g^{-1}\p_t g)(g^{-1}\p_x g)].\label{SU(2)_1}\ee
In the above $g\in$SU(2). Because the homotopy group from $T^2$ (the space-time manifold) to SU(2) is trivial, we can always extend the image $g(x,t)$ to $g(x,t,u)$ so that the former ($g(x,t)$) is the boundary ($g(x,t,u=1$)) of the latter. For $u=0$ the $g(x,t,u=0)$ configuration is taken to be a constant element of SU(2) so that it has a trivial Berry's phase. Therefore $S_{\rm{WZW}}$ represents the Berry's phase term for $g(x,t,u=1)$. Different extensions are guaranteed to yield the same phase factor $\exp(k~S_{\rm{WZW}})$ as long as $k=$ integer.
\subsection{The domain wall between two SPTs: the edge excitation}
The (translation symmetry breaking) perturbation which corresponds to the dimerization in \Fig{fig1} is given by
\be
\lambda \int dxdt~\Tr(g).\label{rel}\ee
The sign of $\lambda$ determines which SPTs is realized. For $g=\exp(i{\theta/2} \hat{n}\cdot\vec{\s})$ \Eq{rel} reduces to $\lambda \cos{\theta/2}$ which favors $\theta=0$ for $\lambda<0$ and $\theta=2\pi$ for $\lambda>0$. Thus a domain wall confined in $-d\le x\le d$ between the $\lambda>0$ and $\lambda<0$ phases is described by the critical theory
subject to the  boundary condition \be g(x=-d,t)=I,~~{\rm{and}}~~g(x=d,t)=-I.\label{bc}\ee The $g(x,t)$ that minimizes $\Tr(\p_xg^{-1}\p_xg)$ and satisfies \Eq{bc} has the form \be g_{\rm{dw}}(x,t)=\exp[i \frac{2\pi (x+d)}{4d}\hat{n}(t)\cdot\vec{\s}],~~~-d\le x\le d.\ee Upon the $u$-extension
$g_{\rm{dw}}(x,t)\ra g_{\rm{dw}}(x,t,u)=\exp[i \frac{2\pi (x+d)}{4d}\hat{n}(t,u)\cdot\vec{\s}]$, where $\hat{n}(t,u=0)=\hat{n}_0$. Substitute the above result into \Eq{SU(2)_1} and integrate $x$ over $[-d,d]$ we obtain the domain wall action
\be
S_{\rm dw/edge}= \frac{1}{2\gamma_2}\int dt(\p_t\hat{n})^2+\frac{i}{2}\int dt\int_0^1du~\hat{n}\cdot\p_t\hat{n}\times\p_u\hat{n},\label{spin}\ee
which is the action of an isolated spin 1/2, or the theory for the boundary.
From the boundary theory (\Eq{spin}) we can directly read off the bulk field theory for the SPT --- simply replace $u$ by a bulk coordinate and remove the $u=0$ boundary condition:
\be
S_{SPT}=\int dt dx \left\{\frac{1}{2\gamma_3}\sum_{j=x,t}(\p_j\hat{n})^2+\frac{i}{2}~\hat{n}\cdot\p_t\hat{n}\times\p_x\hat{n}\right\}.\label{spin2}\ee

For the purpose of later discussions, it is useful to  rephrase the above discussions in terms of a four-component unit vector ``order parameter'' ($\hat{\Omega}=(\Omega_1,\Omega_2,\Omega_3,\Omega_4)$) where
\be
g=\Omega_4 I -i \sum_{j=1}^3 \Omega_j\s_j.
\ee
In terms of $\hat{\Omega}$ \Eq{SU(2)_1} becomes
\be
S_{\rm critical}=\frac{1}{4\gamma_1}\int dxdt (\partial_j \hat{\Omega})^2+\frac{i}{\pi}\int dxdt\int_0^1du\epsilon^{abcd}\Omega_a\p_u \Omega_b\p_t \Omega_c\p_x \Omega_d.\label{vector}\ee The gap opening perturbation in \Eq{rel} becomes \be\lambda \int dxdt~\Omega_4(x,t).\ee Hence $\lambda<0$ favors $\Omega_4>0$ while $\lambda>0$ favors $\Omega_4<0$.
Physically $\Omega_4$ is the dimer order parameter and $\Omega_{1,2,3}$ are the
Neel order parameters of the spin chain. In terms of $\hat{\Omega}(x,t,u)$ the domain wall profile is given by
\be
\hat{\Omega}_{\rm{dw}}(x,t)=\left(\sin{\frac{\pi(x+d)}{2d}}\hat{n}(t),\cos{\frac{\pi(x+d)}{2d}}\right),~~~-d\le x\le d.\label{domain}\ee Upon the $u$-extension
$\hat{\Omega}_{\rm{dw}}(x,t)\ra \hat{\Omega}_{\rm{dw}}(x,t,u)=\left(\sin{\frac{\pi(x+d)}{2d}}\hat{n}(t,u),\cos{\frac{\pi(x+d)}{2d}}\right)$ where $\hat{n}(t,u=0)=\hat{n}_0$. Substitute this expression into \Eq{vector} and integrate $x$ over $[-d,d]$ we reobtain the domain wall action in \Eq{spin}.

The preceding discussion answers the second question we posed in the introduction, namely, what's the relation between the critical theory and the edge/domain wall excitations. We have shown that the interface theory is the critical theory spatially confined between the trivial and non-trivial SPTs. This is the formal
justification of the physical picture (for boundary states) discussed in the introduction.

\subsection{Spinons - the edge excitation dissolved into the critical bulk}\label{spinon}

\begin{figure}
  \centering
  \begin{subfigure}[b]{0.5\textwidth}
     \centering
     \includegraphics[width=\textwidth]{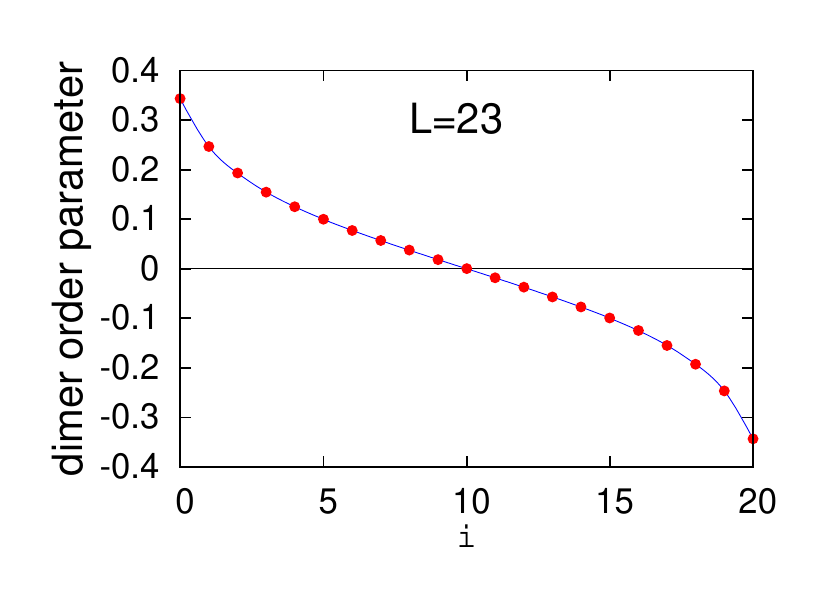}
     \caption{\ }
     \label{fig:dimer23}
  \end{subfigure}%
  ~ 
  \begin{subfigure}[b]{0.5\textwidth}
     \centering
     \includegraphics[width=\textwidth]{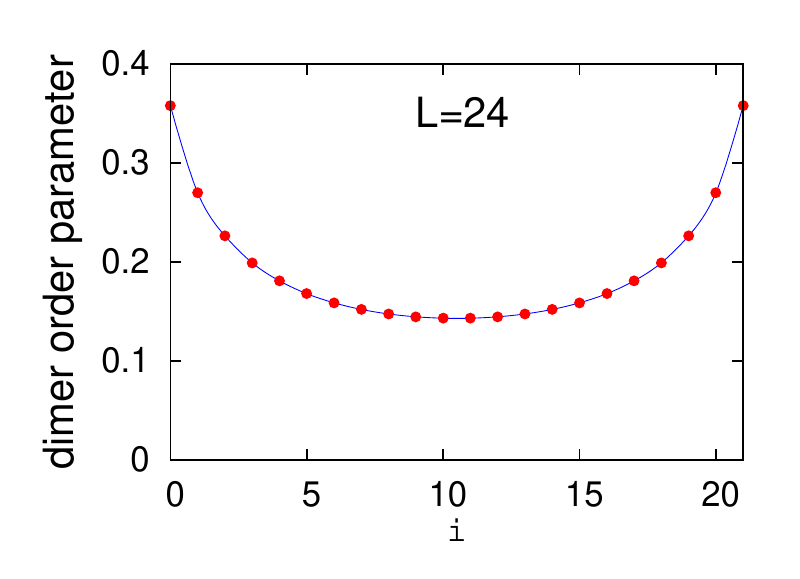}
     \caption{\ }
     \label{fig:dimer24}
  \end{subfigure}
\caption{(color on-line) The dimer order parameter associated with spin 1/2 Heisenberg model under open boundary condition. (a) chain length L=23, (b) chain length L=24.}
\label{dimer}
\end{figure}

In this subsection we address the first question ``what differentiates the critical point of SPTs phase transition from symmetry breaking critical points?''
To get some hints we first look at the results of exact diagonalization for a finite spin 1/2 Heisenberg chain under open boundary condition. The left panel of \Fig{dimer} is the dimer order parameter
\be
(-1)^i\left(\<\vec{S}_i\cdot\vec{S}_{i+1}\>-\<\vec{S}_{i+1}\cdot\vec{S}_{i+2}\>\right)
\ee
for an odd-site chain (L=23) and the right panel is the dimer order parameter for an even-site chain (L=24). Two things are noteworthy. (1) Near the ends of the chain the dimer order parameter is enhanced. This is caused by the breaking of translation symmetry by the ends of the chain. (2) The dimer order parameter for odd/even chain is antisymmetric/symmetric.

In field theory the ends of the chain introduce a local $\lambda_{\rm end}\Omega_4$ perturbation which serves to pin $\Omega_4$. Depending on whether the chain length is even/odd,
the coefficient $\lambda_{\rm end}$ is either of the same or opposite sign, 
which induces the boundary dimer order parameter accordingly. For odd-site chains the $\hat{\Omega}$ field consistent with the boundary condition is
\be
\hat{\Omega}_{\rm{odd}}(x,t,u)=\left(\sin{\frac{\pi x}{L}}\hat{n}(t,u),\cos{\frac{\pi x}{L}}\right).\label{open}\ee Here $L$ is the length of the chain. Substituting the above expression into the WZW term of \Eq{vector} and integrate $x$ over $[0,L]$ we obtain \Eq{spin}, namely the Berry phase of a spin 1/2. This is not surprising because \Eq{open} is the domain wall configuration extended to the entire system!
This ``spinon'' excitation is delocalized throughout the bulk and it is consistent with the ground state of an odd-site chain being $S=1/2$.
Of course when the chain consists of an even number of sites the appropriate field configuration is
\be
\hat{\Omega}_{\rm{even}}(x,t,u)=\left(\sin{\frac{2 k\pi x}{L}}\hat{n}(t,u),\cos{\frac{2 k\pi x}{L}}\right),\label{open1}\ee where $k$ is an integer. In this case the resulting Berry phase we obtain is that for integer spins, which is consistent with the fact that the ground state and excited states of a even-site chain must have integral total spin. 

\begin{figure}[!htb]
\begin{center}
\includegraphics[scale=0.6]
{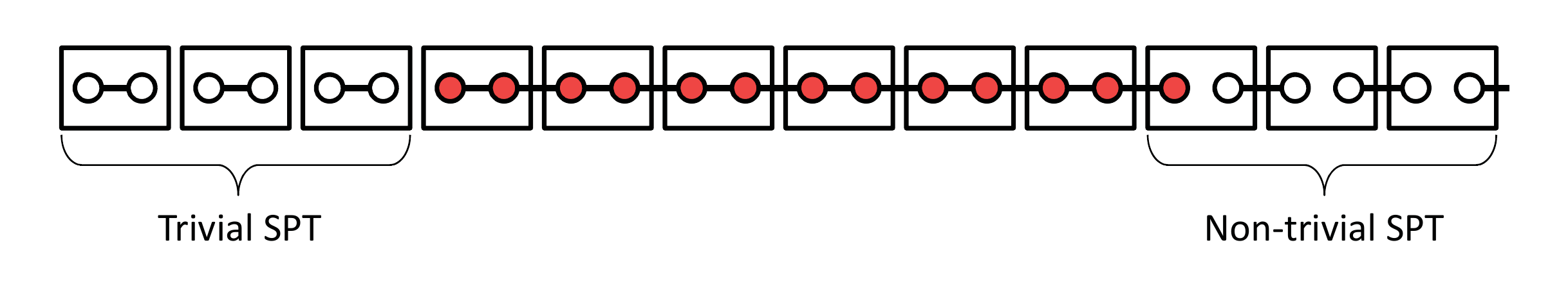}
\caption{(color on-line) An odd-site Heisenberg chain (red dots) is realized as an extended domain wall between the trivial and non-trivial SPT.\label{fig6}}
\end{center}

\end{figure}

Since the very notion of SO(3) SPT requires each unit cell of the system to have integral total spin, the following question immediately arises. Can an odd-site Heisenberg chain be realized in a system whose unit cell has integral spin? While the critical state at phase transition between two SPTs necessarily has integral total spin, it is possible to have an odd-site Heisenberg chain at an extended spatial domain wall between the SPTs, as illustrated in \Fig{fig6}. The rectangular boxes enclose the unit cells and the red dots mark the odd-site Heisenberg chain. Of course what makes this possible is the fact that the bond assignment is that of an extended domain wall between the trivial and non-trivial SPT.

In the literature it is known that the SU(2)$_1$ WZW theory can be viewed as describing a {\it non-interacting gas of spinons} obeying  ``semion exclusion principle''.\cite{ludwig} For even-site chains, spinons must appear in pairs. However for odd-site chains there must be an odd number of spinons. The above discussion clearly suggests that under open boundary condition with suitable chain length, the critical points of SPT phase transitions possess bulk excitations which are the delocalized version of edge excitations. We think this correspondence between the edge and the critical bulk excitations is a special property of the critical points of SPT transitions.  Although we reach the above conclusion by considering a continuous symmetry group, we expect the following conclusions to apply more generally, in particular, for discrete symmetry groups as well : (a) Under the ``domain wall boundary condition'' the critical point of SPT phase transitions possess the delocalized version of the edge excitations; (2) The edge theory is the critical theory confined spatially between two different SPT phases.

\section{The SU(2) SPTs and their phase transition in 2+1 D}
\label{SU(2)_2d}

In 2+1 space-time dimensions the symmetry group SU(2) allows for non-trivial SPTs. Each of them corresponds to an element of $H^3(SU(2),U(1))$ ($=Z$)\cite{xie}. The edge states of these SPTs are governed by the SU(2)$_k$ ($k\in Z$) WZW theory\cite{zxLiu}. Let's start by constructing the $k=1$ SPT, which corresponds to the generator of $H^3(SU(2),U(1))$.

\subsection{The coupled wires construction of the SPTs}

In complete analogy to the previous section, we couple the 1+1 D SU(2)$_1$ WZW edge theories together and gap them. First we say a few words about the edge theory, the 1+1 D SU(2)$_1$ WZW theory. This conformal field theory has SU(2)$_R\times$SU(2)$_L$ symmetry. Viewed as a critical theory in 1+1 D, fine tuning is required to achieve the enlarged $R\times L$ symmetry. On the other hand as the edge theory of a 2+1 D SPT is always gapless and no fine tuning is required. Therefore at low energies the on site SU(2) symmetry must be realized as independent SU(2)$_R\times$SU(2)$_L$ in SPTs. Put it another way, the symmetry that's protecting the gapless edge is the relative SU(2) transformation between the right and left movers, or equivalently an SU(2) symmetry that only acts on either the right or the left movers.\cite{zxLiu}
In the rest of this section, ``SU(2) symmetry'' always refers to this type of relative symmetry transformations.
\begin{figure}[!htb]
\begin{center}
\includegraphics[scale=0.6]
{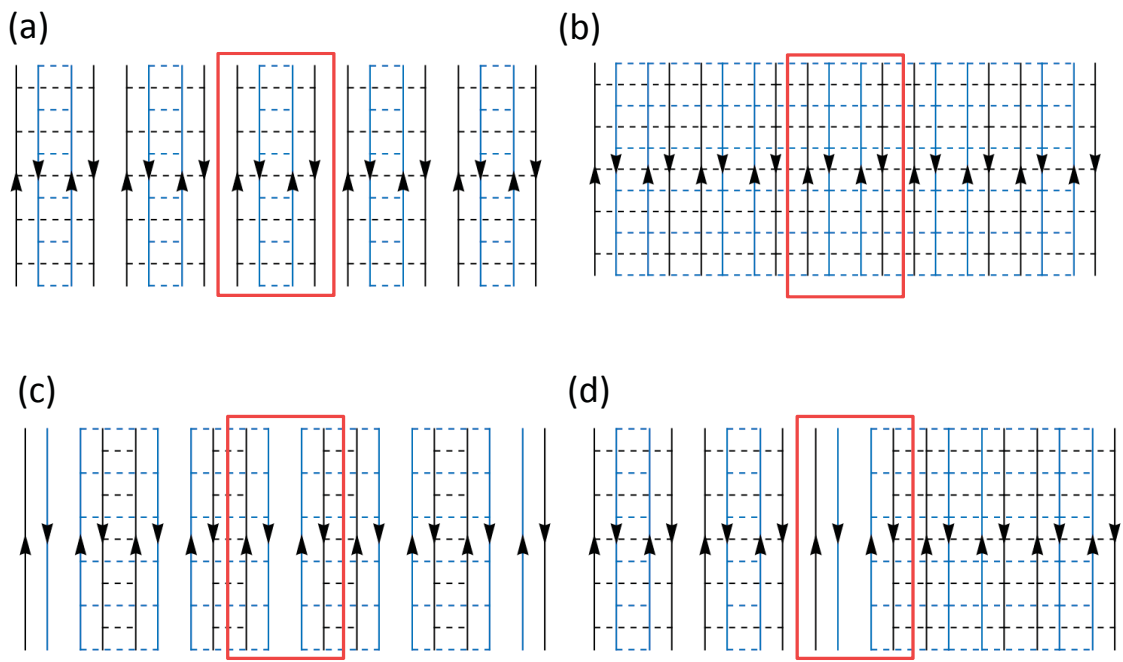}\hspace{0.1in}
\caption{(color on-line) Two 2+1 dimensional SU(2) SPTs and their phase transition. Each pair of up (black) and down (blue) arrowed lines represent a SU(2)$_1$ WZW edge theory. Each unit cell (the red box) encloses two sets of edge modes. There are two set of independent SU(2) transformations, one (SU(2)$_R$) acts on the black modes and the other (SU(2)$_L$) on the blue modes. The black (blue) dashed lines represents the relevant interaction that gaps out the black (blue) edge modes. Due to the SU(2)$_R\times$SU(2)$_L$ symmetry there is no gapping interaction between the black and blue lines. (a) The trivial SPT, (b) the critical point, (c) the non-trivial SPT, (d) the domain wall between (a) and (c).\label{fig2}}
\end{center}
\end{figure}

\Fig{fig2} illustrates the coupled wire construction of the SU(2) SPTs. Each pair of up-down arrowed lines represent a SU(2)$_1$ WZW edge theory (the arrows represent the right and left movers). The SU(2) symmetry  acts only on the black lines (hence on either the right or the left movers). The unit cell is depicted by the red rectangular box. It contains two sets of edge modes and the SU(2) symmetry acts {\it non chirally} because there are both right and left moving black lines in each unit cell. This is analogous to \Fig{fig1} where each unit cell contains {\it integral} spins. Panels (a) and (c) of \Fig{fig2} represent two alternative ways to ``dimerize'' the internal edge states and hence gapping out the bulk excitations. They lead to a trivial and a non-trivial SPTs respectively. In the non-trivial SPT, despite the fact that SU(2) symmetry acts non chirally in each unit cell, on the ``dangling'' edge SU(2) acts chirally (\Fig{fig2}). This is analogous to \Fig{fig1}(c) where, despite the total spin in each unit cell being integral,  at the edges there are unpaired spin 1/2s.

\subsection{The critical theory}
As discussed in the last section, \Eq{vector} is an alternative representation of the SU(2)$_1$ WZW theory. The bulk critical theory (\Fig{fig2}(c)) between the 2+1 D SU(2) SPTs is a WZW vector nonlinear sigma model, where the order parameter has one extra component, in one space dimension higher:\cite{Senthil-Fisher}
\be
\begin{array}{lll}
\tilde{S}_{\rm critical}&=& \frac{1}{2\gamma_4}\int dxdt \sum_{j=x,y,t}(\partial_j \hat{M})^2 + \\
& &\frac{i 3}{4\pi}\int dxdydt\int_0^1du\epsilon^{abcde}M_a\p_u M_b\p_t M_c\p_x M_d\p_y M_e.
\end{array}
\label{vector2}
\ee 
Here $\hat{M}$ is a five component unit vector. The two opposite dimerizations in \Fig{fig2} correspond to $M_5$ havinging opposite signs. The relation between \Eq{vector2} and \Eq{vector} is in exact analogy to the relation between the three-component 0+1 D edge WZW theory (\Eq{spin}) and the four-component 1+1 D critical WZW theory, \Eq{vector}, in the last section.

\subsection{The domain wall theory}
The two SPTs in \Fig{fig2}(a) and (c) can be obtained from \Eq{vector2} by adding the following relevant perturbation 
\be
\lambda \int dt dx dy M_5(x,y,t).
\ee 
$\lambda>0$ and $\lambda<0$ result in the two opposite dimerizations in \Fig{fig2}. Similar to the last section, the edge theory can be obtained by imposing the following boundary condition on the bulk critical theory
\be
M_5(y=-d)=+1 ~~{\rm and}~~M_5(y=d)=-1.\label{bc2}\ee
The $\hat{M}$ profile in the domain wall (spanning $-d\le y\le d$) which minimizes $(\p_y\hat{M})^2$ and satisfies \Eq{bc2}
is given by
\be
\hat{M}_{\rm{dw}}(x,y,t)=\left(\sin{\frac{\pi(y+d)}{2d}}\hat{\Omega}(x,t),\cos{\frac{\pi(y+d)}{2d}}\right),~~~-d\le y\le d\label{domain2}\ee where $\hat{\Omega}$ is
the four-component vector in the edge theory.
Substituting \Eq{domain2} into \Eq{vector2} and integrating $y$ over $[-d,d]$ we obtain
\be
\tilde{S}_{\rm dw/edge}=\frac{1}{2\gamma_5}\int dxdt \sum_{j=x,t} (\partial_{j} \hat{\Omega})^2+\frac{i}{\pi}\int dxdt\int_0^1du\epsilon^{abcd}\Omega_a\p_u \Omega_b\p_t \Omega_c\p_x \Omega_d,\label{edg}\ee namely, the SU(2)$_1$ edge theory. From this edge theory we can directly read off the bulk theory for the SPT phase by replacing $u$ with a spatial coordinate and removing the boundary condition at $u=0$
\be
\tilde{S}_{\rm SPT}=\int dt dx dy\left\{\sum_{j=x,y,t}\frac{1}{2\gamma_6} (\partial_{j} \hat{\Omega})^2+\frac{i}{\pi}\epsilon^{abcd}\Omega_a\p_t \Omega_b\p_x \Omega_c\p_y \Omega_d\right\}.\ee

Therefore again, the  edge theory is the critical theory confined spatially between two different SPTs.

\subsection{Critical excitations as delocalized edge excitations}

Analogous to section \ref{spinon}, when the critical theory is subject to open boundary condition, the boundary introduces a $\lambda_{\rm edge}M_5$ perturbation. When the number of rows in the open, say, $x$ spatial direction is odd, $\lambda_{\rm edge}$ assumes opposite signs on the opposite edges. Under such condition the induced dimer order parameter $M_5$ has opposite signs on the two boundaries. The
$\hat{M}$ profile that minimizes $(\p_x\hat{M})^2$ and matches the above boundary condition has the following form
\be
\hat{M}(x,y,t,u)=\left(\sin{\frac{\pi x}{L}} \hat{\Omega}(y,t,u),\cos{\frac{\pi x}{L}} \right),\ee where $L$ is the linear dimension in the $x$ direction.   Substituting this expression into the WZW term of \Eq{vector2} and integrate $x$ over $[0,L]$ we obtain the edge theory, i.e., \Eq{edg}. Therefore the critical theory \Eq{vector2} with proper boundary conditions describes bulk excitations which corresponds to the edge excitations delocalized in the $x$ direction. Thus the two conclusions reached at the end of the last section are replicated here.

\section{The ``stacked layer construction'' }
\label{stacked_layer}

\begin{figure}[!htb]
\begin{center}
\includegraphics[scale=0.3]
{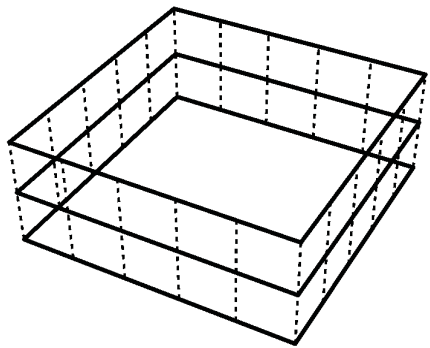}\hspace{0.1in}
\caption{(color on-line) Stack layer construction of SPTs. Before the interlayer coupling, there is a SU(2)$_1$ gapless edge in each layer. \label{fig3}}
\end{center}
\end{figure}

The SU(2) symmetric non-trivial SPT state we obtained in the last section corresponds to the ``generator'' of $H^3(SU(2),U(1))$. Other SPTs can be obtained by ``stacking'' the generator together (\Fig{fig3}). Since all low energy excitations appear on the edge, it suffices to describe how the gapless edges stack together to form new edge excitations. As discussed in Ref.\cite{xie,zxLiu} the edge theory of the SU(2) symmetric SPTs are the SU(2)$_k$ WZW theory.

If we simply stack $k$ independent layers together, the total edge have SU(2)$_1\times$ SU(2)$_1\times$ ... $\times$ SU(2)$_1$ conformal symmetry, which is more than the conformal symmetry of the physical edge, namely, SU(2)$_k$. The ``extra'' conformal symmetry are not required hence should be removed. Upon doing so many edge modes are gapped out. 

Formally this can be achieved by starting with the
edge theory of independent layers
\be
\sum_{n=1}^k \frac{1}{2\gamma_7}\int dxdt\{\Tr(\partial_j g_n^{-1}\partial_j g_n)+
\frac{i}{2\pi}\int_0^1du\Tr[(g_n^{-1}\p_u g_n)(g_n^{-1}\p_t g_n)(g_n^{-1}\p_x g_n)]\},
\ee
then add the inter-layer coupling term
\be
-w \sum_{n=1}^{k-1}\int dx dt~\Tr(g_n^{-1}g_{n+1}).\ee
For large positive $w$ this causes the fields associated with different layers to lock together, namely, $g_1=g_2=...=g_k=g$. The resulting low energy theory
\be
\begin{array}{lll}
\tilde{S}_{\rm edge}&=&\frac{k}{2\gamma_7}\int dxdt \Tr(\partial_j g^{-1}\partial_j g)+ \\
& &\frac{i k}{2\pi}\int dxdt\int_0^1du\Tr[(g^{-1}\p_u g)(g^{-1}\p_t g)(g^{-1}\p_x g)],
\end{array}
\label{SU(2)_k}
\ee 
is precisely the desired SU(2)$_k$ WZW theory.

Alternatively we can represent the edge of each independeny layer by a vector non-linear sigma model
\be
\sum_{n=1}^k\left\{\frac{1}{2\gamma_8}\int dx dt \sum_{j=x,t} (\partial_{j} \hat{\Omega}_n)^2+\frac{i}{\pi}\int dxdt\int_0^1du\epsilon^{abcd}\Omega_{na}\p_u \Omega_{nb}\p_t \Omega_{nc}\p_x \Omega_{nd}\right\}.\ee The inter-layer coupling term now becomes
\be
-w \sum_{n=1}^{k-1}\int dx dt ~\hat{\Omega}_n\cdot\hat{\Omega}_{n+1}.\ee  After the locking $\hat{\Omega}_1=...=\hat{\Omega}_k=\hat{\Omega}$, the theory of the total edge is
\be
\tilde{S}_{\rm edge}=\left\{\frac{k}{2\gamma_8}\int dxdt \sum_{j=x,t} (\partial_{j} \hat{\Omega})^2+\frac{i k}{\pi}\int dxdt\int_0^1du\epsilon^{abcd}\Omega_{a}\p_u \Omega_{b}\p_t \Omega_{c}\p_x \Omega_{d}\right\}.
\label{stack}
\ee

Like before, from \Eq{stack} we can read off the bulk theory for SPT:
\be
\tilde{S}_{\rm SPT}=\int dx dy dt\left\{\frac{1}{2\gamma_9}\sum_{j=x,y,t} (\partial_{j} \hat{\Omega})^2+\frac{i k}{\pi}\epsilon^{abcd}\Omega_{a}\p_t \Omega_{b}\p_x \Omega_{c}\p_y \Omega_{d}\right\}.\ee   If a direct continuous transition exists between the $k$th SPT and the trivial phase, the critical theory should be
\be
\begin{array}{lll}
\tilde{S}_{\rm critical}&=&\frac{1}{2\gamma_{10}}\int dxdt \sum_{j=x,y,t}(\partial_j \hat{M})^2+ \\
& &\frac{i 3k}{4\pi}\int dxdydt\int_0^1du\epsilon^{abcde}M_a\p_u M_b\p_t M_c\p_x M_d\p_y M_e.
\end{array}
\ee 
Again it can be shown (we do not repeat the steps here) that the edge theory is the critical theory confined between two different SPTs.

Because the SU(2) and SO(3) groups do not have non-trivial SPT phase in three spatial dimensions ($H^4(SU(2),U(1))=H^4(SO(3),U(1))=0$),  we stop here.

\section{Conclusion}
\label{conclusion}

In this paper we address the following questions: (1) what's special about the critical theories between different SPTs as compared to the usual Ginzburg-Landau-Wilson theories, and (2) what is the relation between the critical theory of SPT phase transitions and the boundary theory of SPT phases. The answers we find are the following: (1) When subject to appropriate boundary condition, the critical theory possesses the delocalized version of the edge excitations. (2) The edge theory is the bulk critical theory spatially confined between two SPTs. Although special examples are studied (e.g. the symmetry group is continuous) we expect the above conclusions to hold in general. We leave the same study for three space dimensions for the future.

\section{Acknowledgments}

We are in debt to Guang-Ming Zhang and Tao Xiang for stimulating discussions at the initial stage of this work. DHL acknowledges the support by the DOE grant
number DE-AC02-05CH11231.

\end{document}